\documentclass[twocolumn,amsmath,aps,prd,preprintnumbers,amssymb,nofootinbib,showpacs,floatfix]{revtex4}
\usepackage{amssymb}
\usepackage{amsmath}
\usepackage{epsfig}
\usepackage{subfigure}
\usepackage{mathrsfs}
\usepackage{longtable}
\usepackage[usenames,dvipsnames]{xcolor}
\usepackage{mathtools}
\usepackage{relsize}
\begin{document}
%Dan's definitions
 \renewcommand{\thefigure}{\arabic{figure}}
\newcommand{\noj}{}

\newcommand{\apjl}{Astrophys. J. Lett.}
\newcommand{\aap}{Astron. Astrophys.}
\newcommand{\apjs}{Astrophys. J. Suppl. Ser.}
\newcommand{\sa}{Sov. Astron. Lett.}.
\newcommand{\jpb}{J. Phys. B.}
\newcommand{\natu}{Nature (London)}
\newcommand{\aaps}{Astron. Astrophys. Supp. Ser.}
\newcommand{\aj}{Astron. J.}
\newcommand{\aas}{Bull. Am. Astron. Soc.}
\newcommand{\mnras}{Mon. Not. R. Astron. Soc.}
\newcommand{\pasp}{Publ. Astron. Soc. Pac.}
\newcommand{\jcap}{JCAP.}
\newcommand{\jmat}{J. Math. Phys.}
\newcommand{\prep}{Phys. Rep.}
\newcommand{\jtep}{Sov. Phys. JETP.}
\newcommand{\plb}{Phys. Lett. B.}
\newcommand{\pla}{Phys. Lett. A.}
\newcommand{\jhep}{Journal of High Energy Physics}

%Doddy's definitions
\newcommand{\be}{\begin{equation}}
\newcommand{\ee}{\end{equation}}
\newcommand{\bea}{\begin{align}}
\newcommand{\eea}{\end{align}}
\newcommand{\lsim}{\mathrel{\hbox{\rlap{\lower.55ex\hbox{$\sim$}} \kern-.3em \raise.4ex \hbox{$<$}}}}
\newcommand{\gsim}{\mathrel{\hbox{\rlap{\lower.55ex\hbox{$\sim$}} \kern-.3em \raise.4ex \hbox{$>$}}}}
\newcommand{\grad}{\ensuremath{\vec{\nabla}}}
\newcommand{\adotoa}{\ensuremath{{\cal H}}} 
\newcommand{\Uc}{\ensuremath{{\cal U}}}
\newcommand{\Vc}{\ensuremath{{\cal V}}}
\newcommand{\Jc}{\ensuremath{{\cal J}}}
\newcommand{\Mc}{\ensuremath{{\cal M}}}

\newcommand{\unit}[1]{\ensuremath{\, \mathrm{#1}}}

%Ewan's definitions
\newcommand{\gb}{\gamma_{\rm b}}
\newcommand{\dx}{\delta x}
\newcommand{\dy}{\delta y}
\newcommand{\dz}{\delta z}
\newcommand{\dr}{\delta r}
\newcommand{\ds}{\delta s}
\newcommand{\dt}{\delta t}
\newcommand{\uns}{\rmunderscore}
\newcommand{\chimin}{\langle \chi \rangle}

%%%%%%%%%%%%% colortext comments %%%%%%%%%%%%%%
\newcommand{\tkDM}[1]{\textcolor{red}{#1}}                     % Doddy
\newcommand{\tkRH}[1]{\textcolor{blue}{#1}}		  % Ewan
\newcommand{\tkPGF}[1]{\textcolor{green}{#1}}                 % Pedro
\newcommand{\tkEJC}[1]{\textcolor{WildStrawberry}{#1}}   % Ed 
%%%%%%%%%%%%%%%%%%%%%%%%%%%%%%%%%%%%%%%%%%

\title{Axiverse cosmology and the energy scale of inflation}

\author{David J. E. Marsh$^{1,2}\footnote{dmarsh@perimeterinstitute.ca} 
$,~Daniel Grin$^{3}$,~Ren\'{e}e Hlozek$^{4}$, and Pedro G. Ferreira$^{5}$}
\affiliation{$^{1}$Perimeter Institute, 31 Caroline St N,  Waterloo, ON, N2L 6B9, Canada}
\affiliation{$^{2}$Rudolf Peierls Centre for Theoretical Physics, University of Oxford, 1 Keble Road, Oxford, OX1 3RH, UK}
\affiliation{$^{3}$School of Natural Sciences, Institute for Advanced
     Study, Princeton, NJ 08540, USA}
\affiliation{$^{4}$Department of Astronomy, Princeton University, Princeton, NJ 08544, USA}
      \affiliation{$^{5}$Astrophysics, University of Oxford, DWB, Keble Road, Oxford, OX1 3RH, UK}
\date{\today}

 %---------------------- ABSTRACT -------------------------
\begin{abstract}
Ultra-light axions ($m_a\lesssim 10^{-18}$eV), motivated by string theory, can be a powerful probe of the energy scale of inflation. In contrast to heavier axions the isocurvature modes in the ultra-light axions can coexist with observable gravitational waves. Here it is shown that large scale structure constraints severely limit the parameter space for axion mass, density fraction and isocurvature amplitude. It is also shown that radically different CMB observables for the ultra-light axion isocurvature mode additionally reduce this space. The results of a new, accurate and efficient method to calculate this isocurvature power spectrum are presented, and can be used to constrain ultra-light axions and inflation.\end{abstract}

\maketitle

\emph{Introduction}-- Axions \cite{weinberg_comb} are a leading candidate for the dark matter (DM) component of the Universe. Proposed to solve the strong $\mathcal{CP}$ problem, they are also generic in string theory \cite{witten_comb}, leading to the idea of an \emph{axiverse} \cite{axiverse2009}. The number of axions in the axiverse is expected to be large. Due to the topological complexity of string compactifications, and due to non-perturbative physics/moduli stabilisation, the resulting spectrum of axions can cover many decades in mass. Realisations of the axiverse have been achieved in Type-IIB \cite{cicoli2012c} and M-theory \cite{acharya2010a} moduli stabilisation. Beyond the axiverse scenario there are many proposed extensions to the standard model of particle physics (both within string theory and outside of it) that yield new light particles, such as hidden $U(1)$ sectors, minicharged particles, Kaluza-Klein zero modes, generic pseudo Nambu-Goldstone bosons \cite{ringwald2012}, massive gravitons \cite{hinterbichler2011}, galileons \cite{derahm2012}, chameleons \cite{khoury2004}, and axion-like particles. \\
\indent
There are a variety of experimental and observational techniques to search for such particles \cite{ringwald2012}, such as light shining through walls experiments, constraints to fifth forces, stellar cooling, blazar spectra, helioscopes, and black hole super-radiance. Indeed, the population statistics of observed supermassive black holes exclude the existence of light scalar particles in the mass range $10^{-20}\text{ eV}\lesssim m\lesssim 10^{-17}\text{ eV}$ \cite{arvanitaki2010}.
%In the axiverse the axion mass is given by $m_{a}^{2}=\Lambda_{a}^{4}/f_{a}^{2}$. The scales, $f_a$ and $\Lambda_a$, are both determined separately for each axion, and depend on the action, $S$, due to non-perturbative physics on the corresponding cycle: $f_a \sim M_{pl}/S,\Lambda_a^4 = \mu^4 e^{-S}$,
%where $M_{pl}$ is the reduced Planck mass, $M_{pl}^2=1/8\pi G$. The hard non-perturbative scale is $\mu$ and its value should be roughly given by the geometric mean of the Planck scale and the SUSY scale. Solving the strong $\mathcal{CP}$ problem with one of the string axions requires $S\gtrsim 200$ \cite{witten_comb,axiverse2009}, giving rise to stringy values of $f_a \gtrsim 10^{12}\unit{GeV}$, near the GUT scale. The exact value of $S$, however, scales with the area of the corresponding cycle (a dynamically distributed quantity in the landscape), so that small variations in the area lead to exponential variations in the scale of the potential. Thus the mass also depends exponentially on the modulus determining $S$, which is evenly distributed, and so the axion mass spectrum can be taken as an flat distribution on a logarithmic scale \cite{axiverse2009}.
\begin{figure}[htbp!]
\begin{center}
$\begin{array}{@{\hspace{-0.2in}}l}
\includegraphics[width=3.50in, trim=0mm 0mm 20mm 10mm, clip]{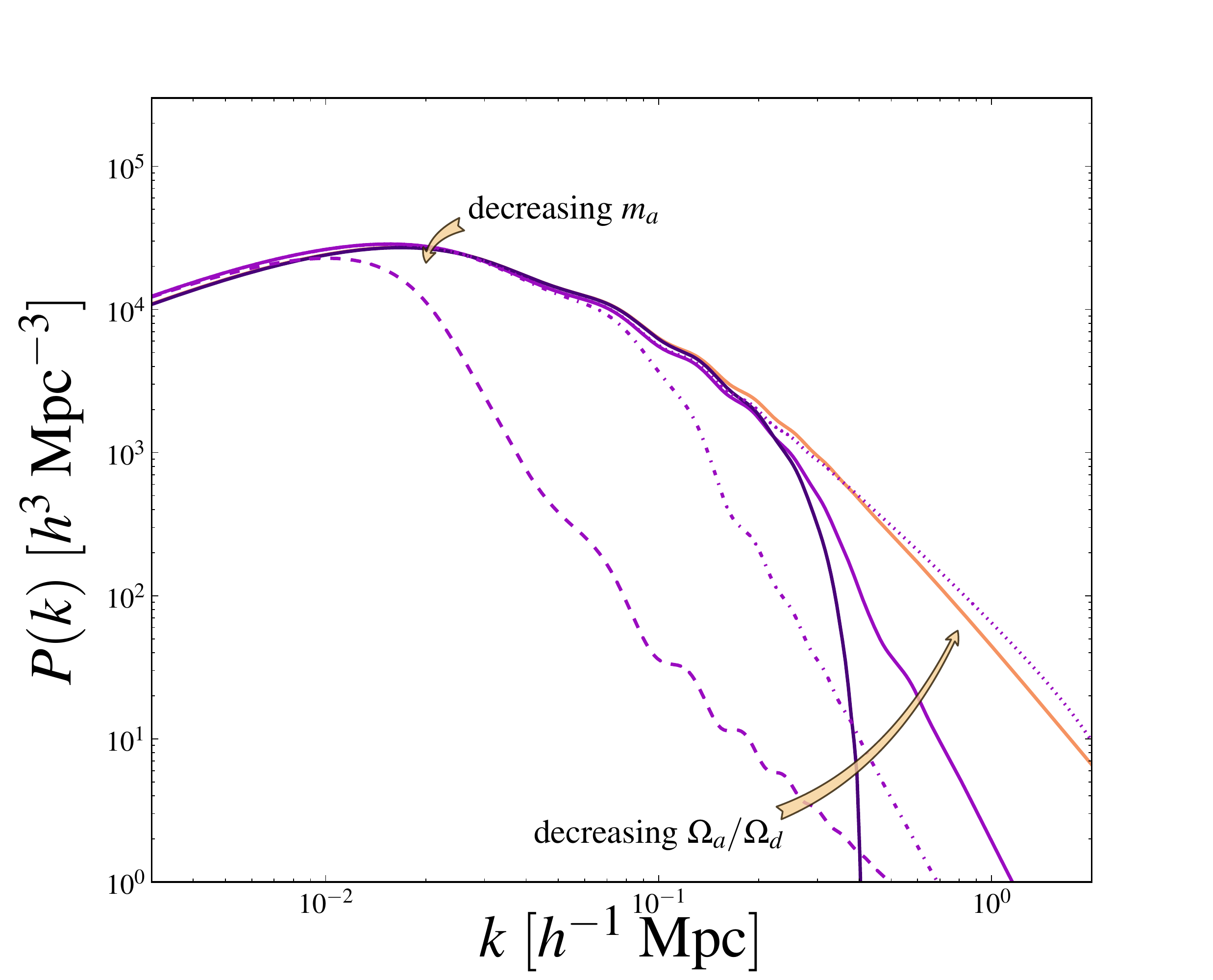}
\end{array}$
\vspace{-0.2in}
\end{center}
\caption{Adiabatic matter power spectra, with varying axion mass $m_a=10^{-28},10^{-26},10^{-25},10^{-23}\text{ eV}$ at fixed density fraction $\Omega_a/\Omega_d=0.5$ (dashed), and varying $\Omega_a/\Omega_d=0.1,0.5,1$ at fixed $m_a=10^{-25}\text{ eV}$ (solid). Spectra are calculated using the methods of Ref.~\cite{marsh/etal:prep}. \label{matterspectra_alldm}}
\end{figure}

At lower masses cosmological observations become increasingly powerful, provided these particles contribute to the energy density of the universe, as DM or dark energy \cite{coupled_de}. For the duration of this paper we will refer solely to axions, though our techniques and results apply to any light particles produced in the same way, and that exist and are massless during inflation. For $m\lesssim 10^{-2}\text{ eV}$ the axion relic density results from vacuum realignment \cite{turner_comb}. An important distinction between QCD axions and lighter axions is that the temperature dependence of the mass drops out and this changes the scalings between misalignment angle and relic density.

If axions are very light, with mass $m_a\lesssim 10^{-18}\unit{eV}$ (ultra-light axions, or ULAs), coherent oscillations of the field lead to the suppression of clustering power on small (but cosmological) scales, and distinguish ULAs from cold (C)DM \cite{hu2000,amendola2005,marsh_comb}. Heuristically, the scale at which structure suppression sets in is the geometric mean of the axion de Broglie wavelength and the Hubble scale. Depending on the axion mass, this scale can affect observed CMB anisotropies and galaxy clustering and weak lensing power spectra. For the classic QCD axion ($m_{a}\sim 10^{-6}\to 10^{-10}~{\rm eV}$), this scale is not cosmologically relevant. In the WKB approximation (averaging over the fast time scale associated with the axion mass, $m_a^{-1}$), the axion may be accurately treated as a fluid, with sound speed
\begin{eqnarray}
c_a^{2}=\left\{\begin{array}{ll}
\frac{k^2}{4m_a^2a^2}&\mbox{if $k\ll 2m_{a}a$},\\
1&\mbox{if $k\gg 2m_{a}a$}.\end{array}\right.\label{heuristic_cs}\end{eqnarray}
The scale of structure suppression begins at the scale $k_{\rm m}$, defined such that those modes with $k>k_{\rm m}$ had sound speed $c_a^2=1$ for some time while they were inside the horizon \cite{marsh_comb}. The effect saturates at the smaller scale $k_{\rm J}=\sqrt{m_{a}H}$. Therefore, like massive standard model neutrinos or more novel warm (W)DM candidates \cite{primackreview}, axions exhibit suppressed structure on small scales, as shown in Fig.~\ref{matterspectra_alldm}. 

%The effect has a completely different origin, however, resulting from the macroscopic `wavy' properties of axions, unlike massive neutrinos, which display suppressed structure because of their relativistic free streaming velocity during structure formation. 
Axions in inflationary cosmology carry \emph{isocurvature} fluctuations \cite{axenides1983}, further distinguishing them from thermally produced CDM. The amplitude of these fluctuations is set by the energy scale of inflation and is thus tied to the amplitude of primordial gravitational waves. Axions thus offer an interesting window into the inflationary epoch, the string landscape, and the multiverse \cite{fox_comb}. For a QCD axion, the isocurvature bounds imply that tensor modes are unobservably small: surprisingly this does not happen for ultra-light axions, and we will shortly explain why.

In  past work, these two aspects of axion cosmology -- the suppression of clustering and the existence of isocurvature perturbations -- have been viewed in isolation. Probes for axion-seeded isocurvature have been restricted to the QCD axion, for which structure suppression on small scales is observationally irrelevant \cite{wmap9,fox_comb}. Meanwhile, observations of CMB anisotropies and galaxy clustering place limits to axion-induced structure suppression \cite{amendola2005}, but do not yet include the isocurvature constraint.  
%The constraints on the axion isocurvature amplitude, when interpreted as constraints on the energy scale of inflation, are degenerate with $f_{ax}=\Omega_a/\Omega_d$, which is also a key parameter related to possible fine tuning in the axiverse. The cosmological, structure-suppressing imprint of ULAs can break this degeneracy between axions and CDM, thus directly check if limits to or measurements of axion isocurvature should be interpreted as limits to the inflationary energy scale, or alternatively, as limits to the ULA mass fraction. As this effect is gravitational in origin, it is independent of axion couplings.

\emph{Isocurvature perturbations, gravitational waves and the CMB}-- It is well known that the tensor-to-scalar ratio, $r$, is a probe of the inflationary energy scale and may be measured using CMB B-modes \cite{kamion_polarization}. The isocurvature amplitude of axions is directly related to $r$, because both ULAs and gravitons are massless during inflation. The standard formulae for the tensor, $\mathcal{P}_h$, and scalar, $\mathcal{P}_\mathcal{R}$, power give the well known result for the tensor-to-scalar ratio $r\equiv \mathcal{P}_{h}/\mathcal{P}_{\mathcal{R}}=16 \epsilon$, where $\epsilon$ is a slow roll parameter.
Given that the scalar amplitude, $A_s=(1/2\epsilon)(H_I/2\pi M_{pl})^2$, is well measured, a measurement of $\epsilon$ constitutes a measurement of the Hubble scale during inflation, $H_I$. The isocurvature fraction also depends on $\epsilon$: measuring it constrains a \emph{function} of $H_I$ and the axion initial misalignment angle \cite{axenides1983}. To measure $H_I$ using isocurvature one must either constrain or make assumptions about the axion initial misalignment angle. Tensor modes only measure $H_I$ if they have an inflationary origin. The sensitivity to tensor modes and isocurvature improves with results from \emph{Planck} \cite{planck_bluebook}.

Isocurvature perturbations are entropy fluctuations of the form $S_{ij}=(\delta n_{i}/n_{i})-(\delta n_{j}/n_{j})$, where the $\delta n_{i}$ and $n_{i}$ are number density fluctuations and average number densities, respectively, in each species present. Entropy fluctuations arise if there are (nearly) massless spectator fields present during inflation \cite{gordon2000}, and the axion is one example.

Since the axion is an independent quantum field from the inflaton, energetically subdominant during inflation, the axion seeds isocurvature perturbations that are \textit{uncorrelated} with the dominant adiabatic fluctuations. The axion isocurvature fluctuations generated in this manner are unavoidable in any standard inflationary scenario as long as neither the inflationary fluctuations of the axion nor reheating restore the Peccei-Quinn symmetry \cite{fox_comb}. This is often the case for the large, stringy, values of $f_a\gtrsim10^{12}$GeV. 

The de Sitter space quantum fluctuations of the axion field $\phi$, have magnitude\footnote{These fluctuations also set the variance on initial misalignment angle and may alter the axion abundance $\Omega_a$ \cite{fox_comb}. In our mass range of interest this effect is negligible.}:
\begin{equation}
\sqrt{\left \langle \delta\phi^{2}  \right \rangle}= \frac{H_I}{2 \pi} \, .
\label{eqn:iso_amplitude}
\end{equation}
%Since the the axion does not evolve during inflation in this model the power spectrum of axion isocurvature fluctuations is also fixed by inflationary slow-roll consistency relations. One finds that $n_i=1-2\epsilon=n_T+1$, where $n_T$ is the spectrum of tensor perturbations. The axion isocurvature power spectrum has a small red tilt. 

There are constraints (e.g. WMAP9 \cite{wmap9}) on the relative amplitude, $\alpha$, of CDM isocurvature fluctuations defined by:
\begin{equation}
\frac{\alpha}{1-\alpha} \equiv \frac{\mathcal{P}_{\mathcal{S}}(k_0)}{\mathcal{P}_{\mathcal{R}}(k_0)} \, \label{eq:axion_iso},
\end{equation}
where $\mathcal{P}_{\mathcal{S}}$ is the isocurvature primordial power spectrum evaluated at pivot wavenumber $k_0$.

The axion power spectrum is given by:
\begin{align}
\left \langle \delta_{a,i}^2 \right \rangle \approx4 \left \langle \left(  \frac{\delta\phi}{\phi} \right)^2\right \rangle &= \frac{(H_I/M_{pl})^2}{\pi^2 (\phi_{i}/M_{pl})^2} \, , \label{eqn:delta_a_init}\\
\left( \frac{\phi_{i}}{M_{pl}} \right)^2 &\approx \frac{6 H_0^2 \Omega_a}{m_a^2 a_{\rm{osc}}^3} \, . \label{eqn:phi_init}
\end{align}
The initial misalignment angle is $\theta_i=\phi_i/f_a$: it is fixed by the relic density and $a_{\text{osc}}$, which is a function of axion mass defined by $3H(a_\text{osc})=m_a$ \cite{marsh_comb,marsh/etal:prep}. Subsequent to $a_\text{osc}$ the axion redshifts as matter, but displays suppression of structure formation.

Before we can relate $\alpha$ to  $m_a, H_I$ and $\Omega_a$ using Eqs.~(\ref{eqn:delta_a_init})-(\ref{eqn:phi_init}), we must clarify the isocurvature normalisation. The usual CDM isocurvature normal mode is defined by taking $\delta_c=1$ as the the initial amplitude of the CDM overdensity, and normalizing the power spectrum such that $\mathcal{P}_\mathcal{S}=\mathcal{P}_{c},$ where $\mathcal{P}_{c}$ is the power spectrum of the CDM fractional overdensity.
\begin{figure}[htbp!]
\begin{center}
$\begin{array}{@{\hspace{-0.34in}}l}
\includegraphics[scale=0.47, trim=2mm 2mm 3mm 2mm, clip]{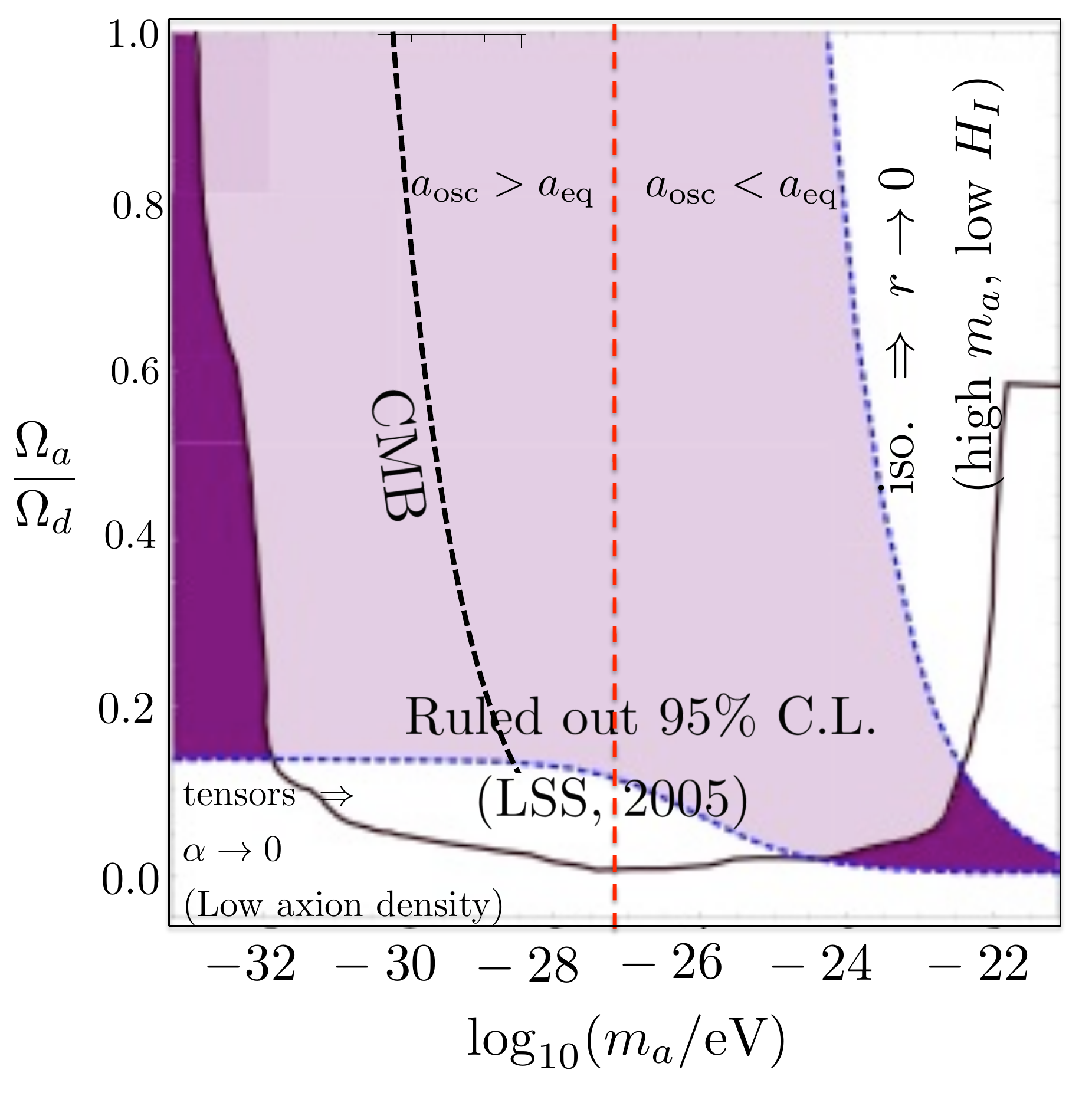}
 \end{array}$
\vspace{-0.25in}
 \end{center}
\caption{Phenomenology in the $\{m_a,\, \Omega_a/\Omega_d\}$ plane. The shaded regions lie between the dashed contours and satisfy $\{0.01<r<0.1,\, 0.01<\alpha_\text{CDM}<0.047\}$, evading current constraints, while being potentially observable with future data. These are not exclusions: outside of the contours either parameter can be large while the other is unobservably small, thereby jointly evading constraints to tensors and isocurvature. The region above the black solid lines, labeled ``Ruled out 95\% C.L. LSS (2005)'', uses the $2\sigma$ adiabatic constraints on $\Omega_a/\Omega_d$ of Ref.~\cite{amendola2005}, and is excluded. The dark shaded regions evade these density constraints yet still have $\{\alpha,\, r\}$ observable so that it may be possible to unambiguously infer $H_I$ from a combination of tensor and isocurvature measurements in the CMB, combined with a LSS measurement of $\Omega_a$. The dashed black line (``CMB'') estimates the modified $\{\alpha,\, r\}$ contours taking into account isocurvature power suppression for low masses (see Fig.~\ref{cmbspectra_isoschematic}).}
\label{fig:constraint_schematic}
\end{figure}

If axions are now included as a sub-component of the DM with the same equation of state and sound speed as CDM (as in Ref.~\cite{wmap9} and others), then there is a single DM effective fluid (with fractional density perturbation $\delta_{d}$) whose isocurvature normal mode is defined by $\delta_d=1$. If axions carry isocurvature fluctuations, while the CDM itself carries only adiabatic fluctuations, then $\mathcal{P}_{\mathcal{S}}=\left(\Omega_{a}/\Omega_{d}\right)^{2}\mathcal{P}_{a}$, where $\mathcal{P}_{a}$ is the axion perturbation power spectrum. In the treatment we develop in Ref.~\cite{marsh/etal:prep}, we incorporate ULAs as a separate effective fluid component with their own independent equation of state and sound speed, in addition to the CDM. In the axion isocurvature normal mode, the initial fractional axion over density is $\delta_{a}=1$, giving $\mathcal{P}_{\mathcal{S}}=\mathcal{P}_{a}$. This yields two definitions of $\alpha$, which we call $\alpha_{\rm CDM}$ (if axions are just included in the overall CDM density) and $\alpha_{a}$ (if axions are treated as a separate species). The WMAP 9-year constraints to axions \cite{wmap9} are derived and stated in terms of $\alpha_{\rm CDM}$.

The two different definitions for the isocurvature fraction are given by
\begin{equation}
\frac{\alpha_a}{1-\alpha_a} =\frac{8 \epsilon}{(\phi_{i}/M_{pl})^2} = \left(\frac{\Omega_d}{\Omega_a} \right)^2 \frac{\alpha_{\rm{CDM}}}{1-\alpha_{\rm{CDM}}} \, .
\label{eqn:alpha_def}
\end{equation}
Measuring the set $\{\alpha,A_s,\Omega_a,m_a\}$ allows one to constrain $H_I/M_{pl}$. For any definition of $\alpha$ one has the well defined prior range $\alpha\in [0,1]$. 

In axion isocurvature models, once $m_a$ and $\Omega_a$ are specified, $r$ (and thus $\epsilon$ and $H_I$) is uniquely determined by $\alpha$, and vice versa. We visualise the interplay of tensor and isocurvature constraints through a schematic plot, Fig.~\ref{fig:constraint_schematic}, which plots contours in $r$ across the entire range of cosmologically relevant $m_a$, using Eqs.~(\ref{eqn:phi_init}), (\ref{eqn:alpha_def}) for a given $\alpha_{\rm{CDM}}$ to fix $\epsilon$ at each point. In particular, when $a_{\rm{osc}}>a_{\rm{eq}}$, where $a_{\text{eq}}$ is the scale factor at matter radiation equality, one finds that $r$ no longer depends on $m_a$ at fixed $\alpha$, and so constraints from ULAs can be markedly different from CDM axions. The two dashed lines span the observable range for $\alpha_\text{CDM}$ and $r$. The isocurvature range is $0.01<\alpha_{\rm{CDM}}<0.047$ where the upper bound is from Ref.~\cite{wmap9} and the lower bound is the forecasted sensitivity of a cosmic variance limited all sky CMB experiment in temperature and polarisation \cite{hamann2009}. The range for tensors is $0.01<r<0.1$ implying a range of sensitivity of an order of magnitude to the energy scale of inflation.

Fig.~\ref{fig:constraint_schematic} shows contours of fixed $r$ and $\alpha$. Areas shaded between these contours have both of observable magnitude. This is in contrast to the QCD axion, due to the different scaling of the relic density, and the very low mass. In the regions of the $\{m_a,\Omega_a/\Omega_d\}$ not shaded by our contours for $r$ and $\alpha$, there are two possibilities: either r or $\alpha$ must be unobservable. If high mass ULAs exist and constitute a sub-leading fraction of the dark matter, bounds to $\alpha$ imply unobservable $r$, and probe low-scale inflation \cite{fox_comb}. Novel to the case of ULAs with $a_\text{osc}>a_\text{eq}$, however, is the fact that if they exist and are energetically important today, existing bounds to the tensor amplitude imply unobservably small $\alpha$. The opposite behaviour comes from the switch in the dependence of the relic density on mass at $a_{\text{osc}}=a_{\text{eq}}$. 

Fig.~\ref{fig:constraint_schematic} also shows the constraints to $\{m_a,\Omega_a/\Omega_d\}$ from Large Scale Structure (LSS) taken from Ref.~\cite{amendola2005}.  Areas of the $\{m_a,\Omega_a/\Omega_d\}$ plane below the contours of Ref.~\cite{amendola2005} are permitted. These constraints severely limit the region where both $r$ and $\alpha$ are simultaneously observable. 

\begin{figure}[htbp!]
\begin{center}
$\begin{array}{@{\hspace{-0.2in}}l}
\includegraphics[scale=0.4]{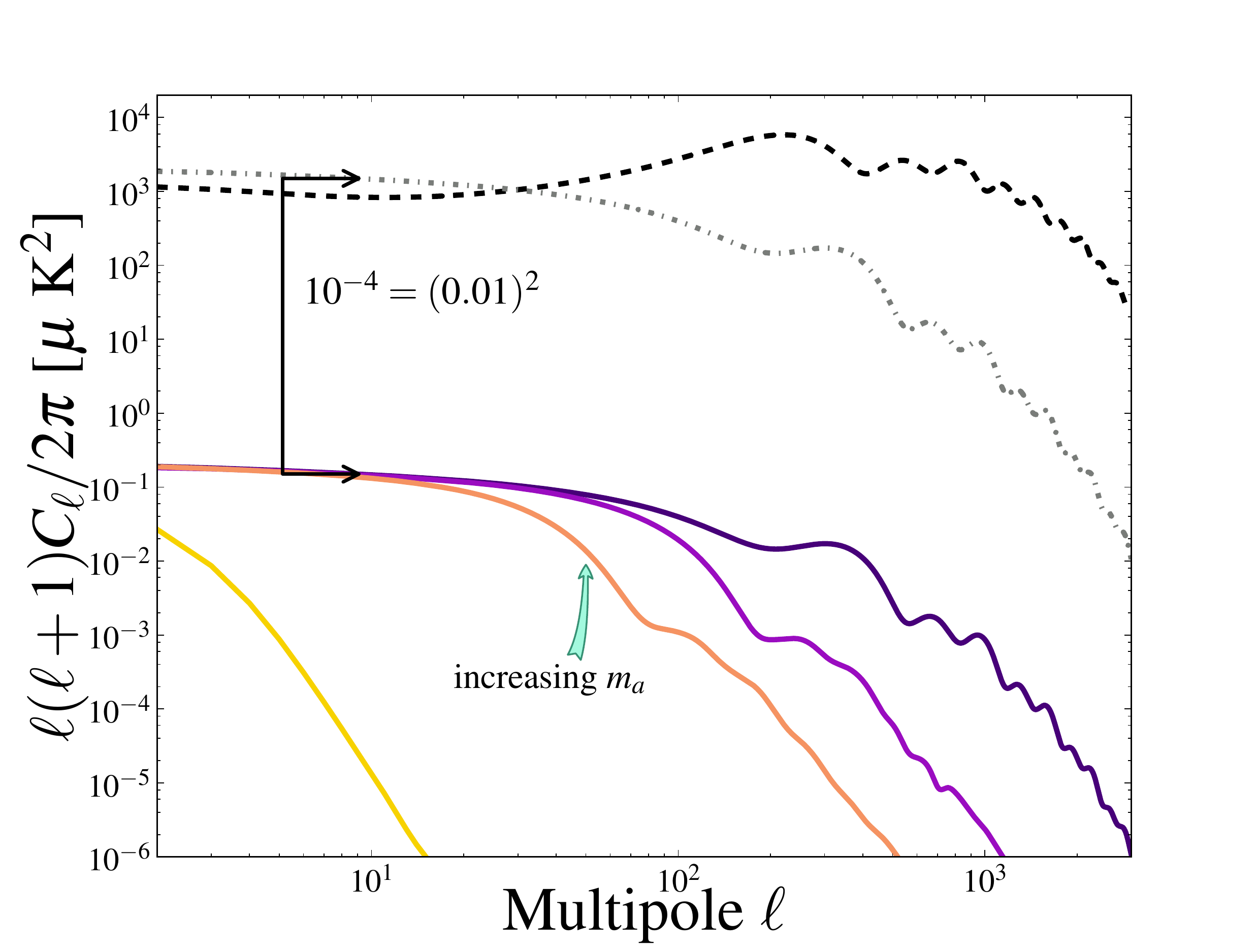}\\ [0.0cm]
 \end{array}$
 \vspace{-0.22in}
\caption{CMB axion isocurvature power spectrum, with adiabatic $\Lambda$CDM for scale (black dashed). We demonstrate the normalisation difference between $\alpha_{\text{CDM}}$ (grey dot-dash) and $\alpha_{a}$ (solid), with $\Omega_a/\Omega_d=0.01$ implying a normalisation difference of $(0.01)^2=10^{-4}$. We also show small-scale power suppression by the lightest axions. The axion masses are $m_a = 10^{-32},10^{-29}, 10^{-28}, 10^{-20} \mathrm{eV}$.}\label{cmbspectra_isoschematic}
\end{center}
\end{figure}

The dark shaded regions in Fig.~\ref{fig:constraint_schematic} are particularly interesting; both regions correspond to simultaneously observable values of $\alpha$ and $r$, while also being consistent with the constraints to $\Omega_a/\Omega_d$ of Ref.~\cite{amendola2005}. Future large scale galaxy redhsift and weak lensing tomography surveys will be able to probe $\Omega_a$ at the sub-percent level \cite{marsh_comb}. In an inflationary context this will break the degeneracy in $\{H_I,\Omega_a\}$, which usually afflicts constraints to $\alpha$. In the shaded regions one can use an $\Omega_a$ detection to predict an observable $\alpha$ from an observed $r$ and vice versa, thus providing a non-trivial cross-check on the inflationary origin of these modes, and thus on $H_I$. Given that there are sources of observable tensor modes possible even with low-scale inflation \cite{senatore2011} these regions provide a novel and truly unambiguous way to measure the energy scale of inflation using the concordance of $\{\alpha,r,\Omega_a\}$. Furthermore, an accompanying isocurvature signal would be strong supporting evidence necessary to infer the axionic origin of any detected suppression of small scale power. We will present constraints in a forthcoming paper \cite{marsh/etal:prep}. Stepping beyond the axiverse paradigm, an isocurvature detection would be evidence that the additional degree of freedom responsible for structure suppression is already present and massless during inflation.

So far we have assumed that constraints to $\alpha_\text{CDM}$ will map over to constraints to $\alpha_a$. For adiabatic fluctuations, the effect of subdominant axions on the CMB observables is very small. For isocurvature fluctuations, however,  the radically different super-horizon solutions \cite{marsh/etal:prep} of axion isocurvature lead to sharply different behavior from the more familiar pure CDM isocurvature. This mode, as well as the more general suppression of small-scale structure in ULA models, is carefully implemented using a modified version of \textsc{camb} \cite{camb} and is described in Ref.~\cite{marsh/etal:prep}. In this case, all other species fall into the gravitational potential wells set up by axions, and so axions drive the behavior of the observables, leading to far more dramatic effects. We show example spectra in Fig.~\ref{cmbspectra_isoschematic}. 

Fig.~\ref{cmbspectra_isoschematic} demonstrates that in the isocurvature mode, CMB power is suppressed on small scales (large $\ell$), with the scale of power suppression becoming larger as the axion mass decreases, just as in $P(k)$ (c.f. Fig.~\ref{matterspectra_alldm}). As the axion mass increases the axion isocurvature spectra asymptote to CDM-like behaviour.

The suppression of power will be important for ULAs in altering the isocurvature constraints. Since the isocurvature power spectrum falls off rapidly at large $\ell$, most constraining power on isocurvature comes from the addition of power along the low-$\ell$ plateau before the first peak at $\ell\sim 200$. When the isocurvature power is suppressed along this plateau the isocurvature spectrum remains significant only at lower and lower $\ell$. Therefore we should expect that not only will allowed values of $\alpha_a$ be different from $\alpha_{\text{CDM}}$ due to normalisation, but also due to the power suppressing properties of ULAs. The effect of this is estimated from the reduced number of modes available to measure isocurvature fraction and is shown in Fig.~\ref{fig:constraint_schematic}. Isocurvature becomes harder to measure and further constrains the observable region for $\{\alpha,r\}$ at the lowest masses, $m_a\lesssim 10^{-28}\text{ eV}$. The lowest mass region is harder to access observationally using LSS measurements since the structure suppressing properties of the axions only occur on very large scales \cite{marsh_comb}. In addition, producing an observable relic density with $m_a\lesssim10^{-28}\text{ eV}$ would require additional physics: for example a large number of axions with nearly degenerate masses.

\emph{Conclusions}-- In this letter we have demonstrated that in the case of ultra-light axions one is able to unambiguously infer the energy scale of inflation from their isocurvature fraction by using large scale structure constraints to bound the relic density. In addition, there are regions of parameter space allowed by current constraints where both the isocurvature fraction and the tensor-to-scalar ratio are within observable reach of near future CMB experiments. This predicted concordance of three observables is a potentially powerful probe of the energy scale of inflation. In the context of the axiverse, the inferred value of $H_I$ from observed tensor modes would predict observable axion isocurvature across more than four orders of magnitude in axion mass. We present constraints to this model in a forthcoming paper \cite{marsh/etal:prep}.
\vspace{-0.25in}
% ------------------------ ACKNOWLEDGEMENTS ----------------------------------
\begin{acknowledgments}
\vspace{-0.1in}
DJEM acknowledges the hospitality of the Department of Astronomy at Princeton University, the Institute for Advanced Study, and the BIPAC, Oxford. DJEM thanks Luca Amendola for providing the results of his previous work, and Cliff Burgess for useful discussions. RH thanks the Perimeter Institute for hospitality. We thank Katie Mack for a careful reading of the manuscript. DG was supported at the Institute for Advanced Study by the National Science Foundation (AST-0807044) and NASA (NNX11AF29G). PGF was supported by STFC, BIPAC and the Oxford Martin School. Research at Perimeter Institute is supported by the Government of Canada through Industry Canada and by the Province of Ontario through the Ministry of Research and Innovation.

\end{acknowledgments}
 
 % ---------------------- BIBLIOGRAPHY -----------------------------------------

%\bibliographystyle{h-physrev3.bst}
\bibliography{doddyoxford,placebib_finf}
\end{document}